\documentclass{lmcs}
\pdfoutput=1

\usepackage{lastpage}
\lmcsdoi{16}{3}{19}
\lmcsheading{}{\pageref{LastPage}}{}{}%
{Jan.~22,~2020}{Sep.~30,~2020}{}

\usepackage[utf8]{inputenc}

\usepackage[english]{babel}
\usepackage{latexsym}
\usepackage{theorem}
\usepackage{url}
\usepackage{xspace}

\usepackage{mathtools}
\usepackage{graphicx}

\usepackage{booktabs}

\usepackage{tikz}
\usepackage{circuitikz}
\usetikzlibrary{decorations.markings,decorations.pathreplacing}
\usetikzlibrary{decorations.pathreplacing}
\usetikzlibrary{arrows.meta}
\usetikzlibrary{intersections}
\usetikzlibrary{bending}
\usetikzlibrary{backgrounds}
\usetikzlibrary{shapes.geometric}
\usetikzlibrary{shapes.misc}
\usetikzlibrary{spy}

{\begin{trivlist}
\item\begin{tabular}{@{}>{\bf}p{2.3em}L@{\ }L@{\ }L@{\ }L@{\ }L@{\ }L@{\ }L@{\ }L}}%
{\end{tabular}\end{trivlist}}

\newcommand{\dataeq}{\approx}
\newcommand{\ndataeq}{\not\approx}
\newcommand{\true}{\mathit{true}}

\newcommand{\Nat}{\mathbb{N}}
\newcommand{\Bool}{\mathbb{B}}

\newcommand{\sort}[1]{\mathit{#1}}
\newcommand{\expr}[1]{\mathit{#1}}

\newcommand{\action}[1]{\mathsf{#1}}
\newcommand{\keyword}[1]{\mathsf{#1}}
\newcommand{\ap}{{:}}
\newcommand{\ignore}[1]{}

\newcommand{\ie}{\emph{i.e.}\xspace}
\newcommand{\eg}{\emph{e.g.}\xspace}
\newcommand{\viz}{\emph{viz.}\xspace}

\newcommand{\None}{{\textit{\hspace{-0.3mm}N\hspace{-0.6mm}o\hspace{-0.3mm}n\hspace{-0.3mm}e}}}

\newcommand{\state}[3]{\draw[thick] (#1,#2) node [shape=circle, inner
    sep=0, minimum size=2.5mm, draw] (#3) {} }
\newcommand{\transition}[6]{\draw[->, thick] (#1) to [out=#5,in=#6] node[#3] {\footnotesize$#2$} (#4)}
\newcommand{\selfloop}[9]{\draw[->, thick] (#1) to [out=#6, in=#7] (#4, #5) node[#3] {\footnotesize$#2$} to [->, out=#8, in=#9] (#1)}


\begin{document}
\title[A symmetric protocol to establish service level agreements]{A symmetric protocol to establish \texorpdfstring{\\}{} service level agreements}
\author[J.F.~Groote]{Jan Friso Groote}
\author[T.A.C.~Willemse]{Tim A.C. Willemse}
\address{Department of Mathematics and Computer Science, Eindhoven University of Technology}
\email{\{J.F.Groote,T.A.C.Willemse\}@tue.nl}
\urladdr{www.win.tue.nl/\~{}\{jfg,timw\}}

\begin{abstract}
  We present a symmetrical protocol to repeatedly negotiate a desired service level between two parties, where
  the service levels are taken from some totally ordered finite domain.
  The agreed service level is selected from levels dynamically proposed by both parties and parties can
  only decrease the desired service level during a negotiation.
  The correctness of the protocol is stated using modal formulas and its behaviour is explained
  using behavioural reductions of the external behaviour modulo weak trace equivalence and divergence-preserving
  branching bisimulation.
  Our protocol originates from an industrial use case
  and it turned out to be remarkably tricky to design correctly.
\end{abstract}

\maketitle

\section{Introduction}

We consider the problem of repeatedly establishing a service level agreement between two parties using a symmetric protocol (\ie, one in which both parties use the same algorithm), without resorting to a centralised decision making process.
The set of service levels is some totally ordered subrange of the natural numbers. The involved
parties independently and dynamically propose desired service levels. When the proposed service levels match, the
protocol must inform both parties which service level is agreed upon. Once a party proposes a service level
this puts an upper bound on it: proposals for lower service levels will still be taken
into account but any subsequent proposal to increase the service level will be ignored.

The problem finds its origin at ASML (\texttt{www.asml.com}), a producer of lithography machines.
In their setting, two (distributed) parties need to determine
whether to swap or flush two chucks holding a silicon wafer. When both parties propose to swap the chucks, swapping
is required. If one party decides to flush, \eg, when a wafer is damaged, the other party has to concede, resulting
in a flush. The swap and flush decisions can be viewed as service levels, with flush the lower of the two service levels.
It turns out it is annoyingly hard to correctly design a symmetric protocol
that solves the problem, symmetry and the absence of a centralised decision making process being part of ASML's initial design requirements for the protocol:
using modal formulas and model checking, all proposed solutions were shown to be incorrect.
Since allowing for a centralised decision making process significantly simplifies the problem, ASML ultimately adopted such a protocol; this solution was subsequently completely
formally modelled and verified~\cite{DBLP:conf/fsen/NeeleRG19}.

This leaves an open question, \viz, whether it is at all possible to have a symmetric protocol to establish a service level agreement between two distributed parties.
Motivated by the many failed attempts 
we set out to prove that no symmetric protocol solving the problem exists.
Much to our surprise, the proof left one scenario open, yielding conditions for a working
negotiation scheme. Generalising the resulting protocol to arbitrary sets of
(ordered) service levels, which has applications in decentralised decision making beyond ASML's restricted setting, turns out to be reasonably straightforward.
We therefore present the more generic repeated service level agreement negotiation protocol for two parties and arbitrary (non-empty) sets of service levels.

The next question is how to provide a convincing argument that our protocol is actually correct.
This is not trivial, because the protocol is continuously interacting with its environment.
As a result, Hoare logics, contracts or other proof techniques for non-reactive systems are not applicable.
Moreover, since the decision making cannot be centralised in a symmetric solution, the two negotiating parties are not necessarily in the same round at every point in time.
This makes it improbable that a solution has an elegant, simple overall external behaviour, implying that it is not feasible to phrase a natural specification of the external behaviour of the protocol and proving conformance of our protocol to that specification.

The only technique that we found suitable to claim the correctness of our solution is to show, by means of model checking, that our formalisation of the protocol meets a number of desirable modal formulas.
These formulas, however, are non-trivial.
Therefore, in addition to verifying the correctness of our protocol, we illustrate aspects of its external behaviour, reduced with respect to weak trace equivalence and divergence-preserving branching bisimulation.
While these pictures are quite insightful for understanding the protocol, they can be quite misleading and cannot be used as correctness arguments; \eg, weak trace equivalence
does not preserve the branching structure of a system.
Moreover, weak trace equivalence masks deadlocks; in fact, we have seen several solutions for which the weak trace equivalence reduced models look perfectly reasonable, but which suffer from deadlocks.

Apart from serving to prove the correctness of our solution, the modal formulas turned out to be instrumental in the process of designing our protocol.
Model checking the formulas
typically quickly revealed logical flaws in variations of our protocol,
and the counter-examples~\cite{DBLP:conf/cade/WesselinkW18,DBLP:conf/csl/CranenLW15} helped to pinpoint
the underlying reasons and steered the process of making the necessary amendments.
It is without a doubt that the use of model checking has greatly accelerated the process of designing our protocol.
\begin{figure}[b]
\begin{center}
\begin{tikzpicture}
\draw[thick] (0,0) rectangle (3,2);
\draw[thick] (4,0.25) rectangle (6,0.75);
\draw[thick] (4,1.25) rectangle (6,1.75);
\draw[thick] (7,0) rectangle (10,2);
\draw[thick, ->] (3,0.5) -- (4,0.5);
\draw[thick, ->] (6,0.5) -- (7,0.5);
\draw[thick, <-] (3,1.5) -- (4,1.5);
\draw[thick, <-] (6,1.5) -- (7,1.5);
\node at (3.5,0.7) {$\action{in_{q}}$};
\node at (6.5,0.7) {$\action{out_{q}}$};
\node at (3.5,1.7) {$\action{out_{q}}$};
\node at (6.5,1.7) {$\action{in_{q}}$};
\draw[thick, ->] (0.6,2) -- (0.6,2.3);
\node at (0.6,2.5) {$\action{agreed}$};
\draw[thick, <-] (2.4,2) -- (2.4,2.3);
\node at (2.4,2.5) {$\action{propose}$};
\draw[thick, ->] (7.6,2) -- (7.6,2.3);
\node at (7.6,2.5) {$\action{agreed}$};
\draw[thick, <-] (9.4,2) -- (9.4,2.3);
\node at (9.4,2.5) {$\action{propose}$};
\end{tikzpicture}
\end{center}
\caption{The structure of two processes negotiating an agreement for quality of service.}%
\label{fig:structure}
\end{figure}
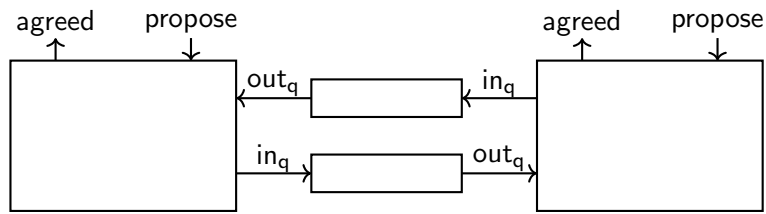
\paragraph{Related Work.} There are various protocols for establishing service level agreements in a variety of domains, see~\eg~\cite{OeyTMOB07,BriquetM06,ParkinKB06}. Most of these protocols focus on asymmetric solutions. Moreover, little effort seems to have been given to formally verifying such protocols. We here focus on the few works we are aware of, and that do use formal verification.

In~\cite{ChaloufKML13}, Chalouf \emph{et al.} study a protocol for service level negotiation, covering
both quality of service and security. With the help of formal
verification using the SPIN model checker, the authors uncover
various deadlocks and propose solutions to fix these deadlocks. Unlike our protocol, which
treats both parties using the protocol symmetrically,
their protocol is based on the \emph{request-response} model, in which one party
initiates negotiation and the other party responds.

The algorithm studied in~\cite{DimitoglouDO06}, which is again asymmetric,
is a bargaining algorithm in which sellers and buyers are to reach an agreement through
a process of `offer/counter-offer'. The algorithm implements a naive negotiation scheme and
terminates upon establishing agreement or
when one party withdraws from the negotiation; repeatedly establishing agreement through the
same algorithm is not supported.  A formal SPIN model is used to simulate the algorithm.

Finally, in~\cite{YaqubYWL12}, Yaqub \emph{et al.} describe a protocol development framework for service level agreement negotiations in cloud and service-based systems.
They illustrate the modelling and verification phase of their framework by developing a multi-tier, multi-round, customisable negotiation protocol called the \emph{Simple Bilateral Negotiation Protocol}.
Also in this case, the protocol is formalised in SPIN\@. The authors abstract from the service levels that can be sent and use a take-it-or-leave-it interaction scheme.
Their formalisation reveals various deadlock and livelock scenarios in early designs of the protocol.
The final model is subjected to a number of LTL properties, focussing mainly on reaching and/or transitioning various phases of the protocol.

We remark that the problem of establishing service level agreements is conceptually close to the (distributed) consensus problem.
Both revolve around reaching agreement on a proposal, but in our case, the proposal can change dynamically in a round; we are not aware of a variation in the distributed consensus problem with similar assumptions.
Distributed consensus is studied for, \eg, systems that communicate (partly) synchronously and asynchronously, and under various types of failures.
In~\cite{FischerLP85}, it is shown that there is no deterministic algorithm to solve the distributed consensus problem in an asynchronous message passing setting, in case there is at least one failing process; we expect that this result extends to our setting when generalising our problem to multiple processes which are allowed to fail.

\paragraph{Outline.}
In the subsequent sections we present the protocol in the process specification language mCRL2 (Section~\ref{sect3}) and
explain its external behaviour, reduced with respect to weak trace equivalence and divergence-preserving branching bisimulation in Section~\ref{sect4}. In Section~\ref{sect5}, we
formulate the requirements as modal formulas and verify them for increasing ranges of qualities of service. In Section~\ref{sect6}, we
state some conclusions and future work. We first phrase the problem statement and the requirements the protocol must satisfy in Section~\ref{sec:requirements}.

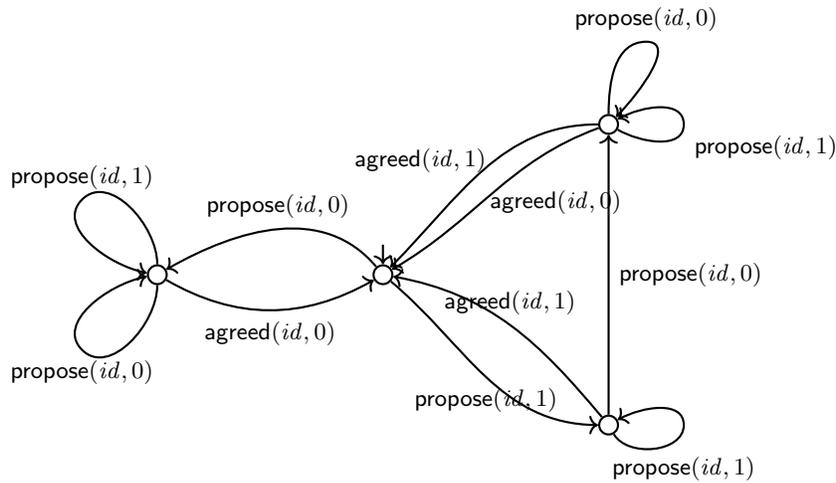
\begin{figure}[t]
\begin{center}
\begin{tikzpicture}
\state{0}{0}{A};
\draw [thick, ->] (0,0.4) -- (A);
\state{-3}{0}{B};
\state{3}{2}{C};
\state{3}{-2}{D};
\transition{A}{\footnotesize\action{propose}(\mathit{id},0)}{above}{B}{130}{30};
\transition{B}{\action{agreed}(\mathit{id},0)}{below}{A}{-30}{210};
\selfloop{B}{\action{propose}(\mathit{id},1)}{above}{-4}{1}{90}{45}{225}{170};
\selfloop{B}{\action{propose}(\mathit{id},0)}{below}{-4}{-1}{-90}{-45}{-225}{-170};

\transition{A}{\action{propose}(\mathit{id},1)}{below}{D}{-40}{180};
\transition{D}{\action{agreed}(\mathit{id},1)}{above}{A}{130}{-10};
\transition{D}{\action{propose}(\mathit{id},0)}{right}{C}{90}{-90};
\transition{C}{\action{agreed}(\mathit{id},1)}{above left=-5}{A}{180}{45};
\transition{C}{\action{agreed}(\mathit{id},0)}{below right=-5}{A}{200}{30};

\selfloop{D}{\action{propose}(\mathit{id},1)}{below=8}{4}{-2}{-60}{-90}{90}{30};
\selfloop{C}{\action{propose}(\mathit{id},1)}{below right}{4}{2}{-30}{-90}{90}{30};
\selfloop{C}{\action{propose}(\mathit{id},0)}{above}{3.5}{3.1}{90}{180}{0}{45};

\end{tikzpicture}
\end{center}
\caption{The external behaviour for one single party $\mathit{id}$ with two qualities of service}%
\label{fig:weak_trace_single_party}
\end{figure}

\begin{figure}
\begin{center}
\begin{tikzpicture}
\draw[rounded corners, thick] (-0.3,0) -- (0,0) -- (0,1) -- (-0.3,1);
\draw (-1.4,0.2) node {$\action{agreed}(\mathit{id}_1,0)$};
\draw (-1.4,0.7) node {$\action{propose}(\mathit{id}_1,0)$};
\draw[rounded corners, thick] (-0.3,1.2) -- (0,1.2) -- (0,3.2) -- (-0.3,3.2);
\draw (-1.4,1.4) node {$\action{agreed}(\mathit{id}_1,0)$};
\draw (-1.4,2.0) node {$\action{propose}(\mathit{id}_1,1)$};
\draw (-1.4,2.8) node {$\action{propose}(\mathit{id}_1,0)$};

\draw[rounded corners, thick] (-0.3,3.4) -- (0,3.4) -- (0,5.6) -- (-0.3,5.6);
\draw (-1.4,3.6) node {$\action{agreed}(\mathit{id}_1,0)$};
\draw (-1.4,4.1) node {$\action{propose}(\mathit{id}_1,1)$};
\draw (-1.4,5.3) node {$\action{propose}(\mathit{id}_1,0)$};

\draw[rounded corners, thick] (-0.3,5.8) -- (0,5.8) -- (0,7.8) -- (-0.3,7.8);
\draw (-1.4,6.1) node {$\action{agreed}(\mathit{id}_1,1)$};
\draw (-1.4,6.6) node {$\action{propose}(\mathit{id}_1,1)$};
\draw (-1.4,7.3) node {$\action{propose}(\mathit{id}_1,1)$};

\draw[rounded corners, thick] (0.5,1.9) -- (0.2,1.9) -- (0.2,0) -- (0.5,0);
\draw (1.6,0.2) node {$\action{agreed}(\mathit{id}_2,0)$};
\draw (1.6,0.9) node {$\action{propose}(\mathit{id}_2,0)$};
\draw (1.6,1.6) node {$\action{propose}(\mathit{id}_2,1)$};

\draw[rounded corners, thick] (0.5,3.7) -- (0.2,3.7) -- (0.2,2.1) -- (0.5,2.1);
\draw (1.6,2.4) node {$\action{agreed}(\mathit{id}_2,0)$};
\draw (1.6,3.3) node {$\action{propose}(\mathit{id}_2,0)$};

\draw[rounded corners, thick] (0.5,3.9) -- (0.2,3.9) -- (0.2,4.6) -- (0.5,4.6);
\draw (1.6,4.075) node {$\action{agreed}(\mathit{id}_2,0)$};
\draw (1.6,4.425) node {$\action{propose}(\mathit{id}_2,0)$};

\draw[rounded corners, thick] (0.5,4.8) -- (0.2,4.8) -- (0.2,7.8) -- (0.5,7.8);
\draw (1.6,5.1) node {$\action{agreed}(\mathit{id}_2,1)$};
\draw (1.6,7.5) node {$\action{propose}(\mathit{id}_2,1)$};

\end{tikzpicture}
\end{center}
\caption{Four rounds of the protocol}%
\label{fig:asynchronous_rounds}
\end{figure}
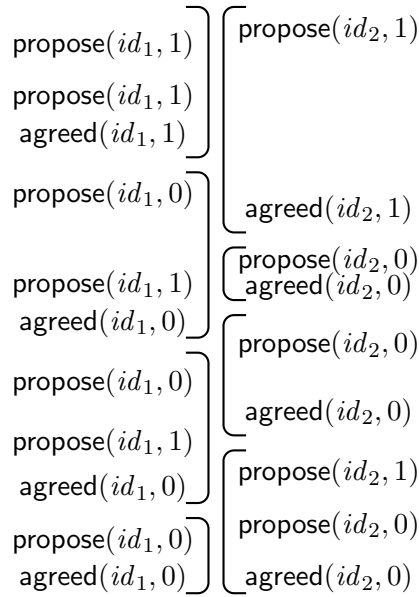
\pagebreak[3] 
\section{Requirements for a service level agreement negotiation protocol}%
\label{sec:requirements}
The problem we address is that of repeatedly, in \emph{rounds}, negotiating a service level agreement between two parties (identified by $\mathit{id}_1$ and $\mathit{id}_2$, respectively), where (1)
the decision making process is decentralised and (2) both parties use an instance of the same algorithm, which is oblivious to which party is running it.
Communication between the two parties proceeds asynchronously via message passing along reliable channels, and the set of service levels is a totally ordered subrange of the natural numbers.
The required architecture is depicted in Figure~\ref{fig:structure}.

In each round, both parties aim to establish a common service level.
Service levels can be proposed using the action
$\action{propose}(\mathit{id},l)$ where $\mathit{id}$ is the identifier of this party and $l$ is a
service level.
A party may repeatedly propose service levels and the protocol is to process all these proposals.
Once a party proposes a service level, subsequent proposals for higher service levels must be ignored but lower service levels may still be agreed upon.
A round ends for this party when the action
$\action{agreed}(\mathit{id},l')$ takes place. The party can then start the negotiation for a new common service level in this new round. The service levels proposed in one round are not to affect the decision on a common service level in a later round.\medskip

From the perspective of a single party $\mathit{id}$, a solution to the problem must meet the following requirements on the agreed quality of service at the end of a round:
\begin{enumerate}
\item[(I)]
The agreed quality of service $l'$ must have been proposed by party $\mathit{id}$ in the same round.
\item[(II)]
If at some time in a round a service level $l$ is proposed by party $\mathit{id}$
, a higher service level $l'$ proposed only after $l$
will never be agreed in this round.
\end{enumerate}
It is allowed that a party successively proposes different services levels $l$ and $l'$ with
$l'{<}l$. In this case the protocol can agree on both service levels.
Figure~\ref{fig:weak_trace_single_party} depicts
a labelled transition system that represents the allowed sequences of proposals and agreements that can occur from
the perspective of one process, if there are only two service levels to choose from.
This figure nicely illustrates that when a proposal for level $1$ is followed by a proposal to settle on level
$0$, both can be agreed, as after doing the actions $\action{propose}(\mathit{id},1)$ followed by
$\action{propose}(\mathit{id},0)$ from the initial state,
it is possible to do both the actions $\action{agreed}(\mathit{id},0)$ and $\action{agreed}(\mathit{id},1)$.
But if initially service level $0$ is proposed by doing action $\action{propose}(\mathit{id},0)$ in the initial state,
service level $0$ is the only service level that can be agreed upon, as $\action{agreed}(\mathit{id},0)$ is the only
agreed action after this proposal.

If we consider two parties simultaneously, we observe that their perspectives on when rounds start and end may not be the same. Figure~\ref{fig:asynchronous_rounds} illustrates such a scenario. We therefore speak of round $n$ for a particular party and not for the protocol as a whole.
Round $1$ for party $\mathit{id}$ is the time period from the start of the protocol up to and including the time of the first
action $\action{agreed}(\mathit{id},l)$. Round $n{+}1$ for party $\mathit{id}$ is the time period just after
the $n^\mathit{th}$ occurrence of an action $\action{agreed}(\mathit{id},l')$ up to and including the $n{+}1^\mathit{th}$ occurrence of
$\action{agreed}(\mathit{id},l'')$.

From the perspective of both parties, the solution provided by the protocol should meet
the following two requirements on the quality of service both parties agree upon:
\begin{enumerate}
\item[(III)]
If during round $n$ for party $\mathit{id}_1$, action $\action{agreed}(\mathit{id}_1,l)$ occurs and during round $n$ for party $\mathit{id}_2$ action
$\action{agreed}(\mathit{id}_2,l')$ happens, then $l=l'$.
\item[(IV)] Whenever during round $n$ both parties propose a common service level $l$ through actions $\action{propose}(\mathit{id}_1,l)$ and $\action{propose}(\mathit{id}_2,l)$, then both
parties will reach an agreement, indicated by $\action{agreed}(\mathit{id}_1,l')$ and
$\action{agreed}(\mathit{id}_2,l')$ for some $l'$ possibly different from $l$.
The execution of the agreement action can be postponed indefinitely only by the parties repeatedly
submitting redundant proposals, or by failing to propose the essential matching $\action{propose}(\mathit{id},l')$,
with $l'$ the minimal service level proposed by the other party and not party $\mathit{id}$ itself.

\end{enumerate}
In case of requirement (IV) there are two typical example traces in which an action $\action{agreed}$
does not need to follow within a finite number
of steps.
The first example scenario is one in which a process executes a $\action{propose}(\mathit{id}_1,1)$ action and subsequently the other process
repeatedly executes action $\action{propose}(\mathit{id}_2,1)$ (all of which get ignored, except for the first one). This effectively prevents the
protocol to report an action $\action{agreed}(\mathit{id}_1,1)$ and $\action{agreed}(\mathit{id}_2,1)$, although
the protocol can in principle do these.

The second one is process $\mathit{id}_1$ --- probably repeatedly ---
executes action $\action{propose}(\mathit{id}_1,1)$ and process $\mathit{id}_2$ executes action
$\action{propose}(\mathit{id}_1,0)$. As they propose different service levels, an agreement never follows.
The protocol ends up in the middle topmost state in Figure~\ref{fig:branchbis} (note that in this figure self-loops
in states have been omitted for readability).

\section{Formal model in mCRL2}%
\label{sect3}
In this section we give a full model of our protocol in the process specification language mCRL2~\cite{GM14}.
Somewhat to our surprise the protocol is rather concise, especially because it also includes the description of
the unbounded channels used by both parties to exchange messages. Being a formal description, it is very straightforward
to translate it to any appropriate implementation language.

\subsection{A note on mCRL2}

A detailed description of the specification language mCRL2 is available in the standard textbook~\cite{GM14}.
To ensure that our presentation of the protocol is self-contained, we here briefly review the operators used in our
specification.
\emph{Actions} are the basic building blocks of a specification. In our protocol, the actions typically represent
activities such as sending or receiving messages.
Using the binary sequential composition operator $\cdot$, the binary non-deterministic choice operator $+$ and
recursion, processes can be built compositionally.
The binary non-deterministic choice is generalised by the quantifier $\sum$, taking a typed variable and a process as arguments.
Intuitively, an expression of the form $\sum x{:}T. p$ represents a non-deterministic choice among all possible processes resulting from substituting concrete values, taken from $T$, for $x$ in $p$.
Recursive specifications can carry typed data parameters and data expressions can be passed as arguments in recursive calls. Finally, the process $b \to p \diamond q$  behaves as $p$ when Boolean expression $b$ evaluates to $\true$ and $q$ otherwise.
Two processes can be composed in parallel using the $||$ operator. A binary communication operator, taking a \emph{communication function}
and a process as arguments, can be used to specify that two (or more) actions that can happen in parallel, synchronise and create a new action.
The binary \emph{allow} operator, taking a set of actions $A$ and a process $p$, can be used to describe the behaviour of $p$ when all actions not in set $A$ are blocked; this operator can be used to `enforce' synchronisation by only allowing the actions resulting from a communication, and blocking the individual actions involved in the communication.

\subsection{A service level agreement negotiation protocol}

We use natural numbers $\Nat$ to represent service levels and the standard ordering $\leq$ on natural numbers to model the total order on the service levels, assuming that $0$ is the lowest level attainable.
The set $\expr{Levels}$, defined as the set $\{\,l\ap \Nat \mid l < \expr{Max}\,\}$, is the set of \emph{valid} levels, where we assume that the value $\expr{Max}> 0$ is a parameter in our model.
The natural number $\None$, which we set to $\expr{Max}$, is used to represent that no service level has been settled yet.

We distinguish between messages used to propose a particular service level, and messages used to communicate that a party decided upon a service level.
This is formalised by the data sort $\sort{Message}$, defined as follows:
\[
\sort{Message} = \keyword{struct}~ \expr{inform}(\Nat) ~|~ \expr{decide}(\Nat)
\]

The two parties are identified by means of an identifier, taken from a finite collection of identifiers $\sort{ID}$:
\[
\sort{ID} = \keyword{struct}~ \expr{id_1} ~|~ \expr{id_2}
\]
Both parties communicate asynchronously via message passing along a uni-directional, reliable channel.
The channel can receive messages (via parameterised action $\action{r_q}$) from one party and pass messages (via parameterised action $\action{s_q}$) to the other party.
Process $Q$, depicted below, models the (first-in first-out) channel for party $\expr{id}$;
the messages sent along the channel are recorded in a queue $\expr{q}$ of messages, modelled as a list.

\[
\begin{array}{l}
\keyword{proc}~Q(\expr{id}\ap\sort{ID},\expr{q}\ap\sort{List(Message)}) = \\
\quad \sum\limits_{m\ap\sort{Message}}~\action{r_q}(\expr{id}, \expr{m})
                                        \cdot Q(\expr{id}, \expr{m \triangleright q})
+ (|q|>0) \rightarrow \action{s_q}(\expr{id},\expr{rhead(q)}) \cdot Q(\expr{id}, \expr{rtail(q)})
\end{array}
\]

Remark that $m \triangleright q$, for a list $q$ and element $m$ represents a new list $q$ with $m$ prepended to it.
The expression $\expr{rhead}(q)$, the \emph{rear head}, denotes the last element in $q$, whereas expression $\expr{rtail}(q)$ denotes the list without its last element. Note that $\expr{head}(q)$ and $\expr{tail}(q)$, used elsewhere, denote the regular head and tail of a queue.
The length of list $q$ is denoted by $|q|$.

We next describe the algorithm (the protocol entity) that each of the negotiating parties runs; see Figure~\ref{fig:protocol} for the mCRL2 specification.
Each entity uses actions $\action{propose}(\mathit{id},l)$ and $\action{agreed}(\mathit{id},l)$
to communicate with its external environment; these actions represent the events explained in the previous section.
Communication \emph{between} the two parties proceeds via two instances of the process $Q$.
Each protocol entity can send messages to this queue via action $\action{s}$ and receive messages from the queue via action $\action{r}$.
The synchronisation between these actions and the actions used by process $Q$ is described by process $\expr{SLAN}$ at the end of this section.

The protocol entity $\expr{P}$ (see the first equation in Figure~\ref{fig:protocol}) is logically decomposed in two distinct processes: a process $\expr{P_{other}}$, handling the messages that arrive from the other party, and a process $\expr{P_{env}}$ that is triggered by communications with the external environment.
Decision making is distributed between these two subprocesses: both proposals for a new service level and messages informing the protocol entity that the other party has proposed a particular service level may lead to decisions being taken.
Note that decisions are made by a protocol entity itself, \ie, without any centralised party or manager.

\begin{figure}[h]
\centering
\[
\boxed{
\begin{array}{l}
\keyword{proc}~\expr{P}(\expr{id},\expr{from}\ap\sort{ID}, \expr{mine}\ap \sort{Set}(\Nat), \expr{theirs} \ap \sort{\Nat}, \expr{decision} \ap \Nat, \expr{hold} \ap \Bool) = \\

\enskip \expr{P_{env}}(\,) + \expr{P_{other}}(\,) \\
\\

\keyword{proc}~\expr{P_{env}}(\expr{id},\expr{from}\ap\sort{ID}, \expr{mine}\ap \sort{Set}(\Nat), \expr{theirs} \ap \sort{\Nat}, \expr{decision} \ap \Nat, \expr{hold} \ap \Bool) = \\
\enskip (\expr{decision} \dataeq \None)~\rightarrow \expr{P_{negotiate}}(\,) \diamond \expr{P_{decide}}(\,) \\
\\
\keyword{proc}~\expr{P_{decide}}(\expr{id},\expr{from}\ap\sort{ID}, \expr{mine}\ap \sort{Set}(\Nat), \expr{theirs} \ap \sort{\Nat}, \expr{decision} \ap \Nat, \expr{hold} \ap \Bool) = \\

\enskip \action{agreed}(\expr{id},\expr{decision}) \cdot \expr{P}(\expr{decision} = \None) + {} \\
\enskip \sum\limits_{l \ap \Nat}~ (l \in \expr{Levels}) \rightarrow \action{propose}(\expr{id},l) \cdot \expr{P}(\,) \\
\\
\keyword{proc}~\expr{P_{negotiate}}(\expr{id},\expr{from}\ap\sort{ID}, \expr{mine}\ap \sort{Set}(\Nat), \expr{theirs} \ap \sort{\Nat}, \expr{decision} \ap \Nat, \expr{hold} \ap \Bool) = \\
\enskip \sum\limits_{l \ap \Nat}~ (l \in \expr{Levels}) \rightarrow \action{propose}(\expr{id},l) \cdot {} \\
\enskip\enskip \enskip \enskip ( l < \expr{min}(\expr{mine})  \wedge l \leq \expr{theirs}) \\
\enskip\enskip \enskip \enskip \enskip \rightarrow~ ( ( l \dataeq \expr{theirs} ) \\
\enskip\enskip \enskip \enskip \enskip \enskip \enskip\enskip \rightarrow \action{s}(\expr{id},\expr{decide}(l)) \cdot \expr{P}(\expr{mine} =\emptyset, \expr{theirs} = \None, \expr{decision} = l, \expr{hold} = \expr{true}) \\
\enskip\enskip \enskip \enskip \enskip \enskip \enskip\enskip \diamond~  \action{s}(\expr{id},\expr{inform}(l)) \cdot \expr{P}(\expr{mine} =\expr{mine} \cup \{l\}) \\
\enskip\enskip \enskip \enskip \enskip \diamond~ \expr{P}(\,) \, ) \\
\\

\keyword{proc}~\expr{P_{other}}(\expr{id},\expr{from}\ap\sort{ID}, \expr{mine}\ap \sort{Set}(\Nat), \expr{theirs} \ap \sort{\Nat}, \expr{decision} \ap \Nat, \expr{hold} \ap \Bool) = \\
\enskip \sum\limits_{l \ap \Nat}~ \expr{hold}
\rightarrow~ ( \action{r}(\expr{from},\expr{inform}(l)) \cdot \expr{P}(\,) +
\action{r}(\expr{from},\expr{decide(l)})\cdot \expr{P}(\expr{hold} = \expr{false})  ) \\
\enskip \enskip \enskip \diamond~ (\, \action{r}(\expr{from},\expr{decide}(l))\cdot \action{s}(\expr{from},\expr{decide}(l)) \cdot \expr{P}(\expr{mine} = \emptyset, \expr{theirs} = \None, \expr{decision} = l) + {} \\
\enskip \enskip \enskip \phantom{\diamond~ (} \action{r}(\expr{from},\expr{inform}(l)) \cdot \\ 
\enskip \enskip \enskip \enskip \enskip\enskip (l \in \expr{mine} \wedge \expr{decision} \dataeq \None) \\
\enskip \enskip \enskip \enskip \enskip\enskip \enskip\enskip \rightarrow~ \action{s}(\expr{id},\expr{decide}(l)) \cdot \expr{P}(\expr{mine} = \emptyset, \expr{theirs} =\None, \expr{decision} = l, \expr{hold} = \expr{true}) \\
\enskip \enskip \enskip \enskip \enskip\enskip \enskip\enskip \diamond~ \expr{P}(\expr{theirs} = \expr{min}(\expr{theirs},l) \, )\,)\,) 
\end{array}
}
\]
\caption{Model of the protocol entity negotiating the desired service level agreement.}%
\label{fig:protocol}
\end{figure}
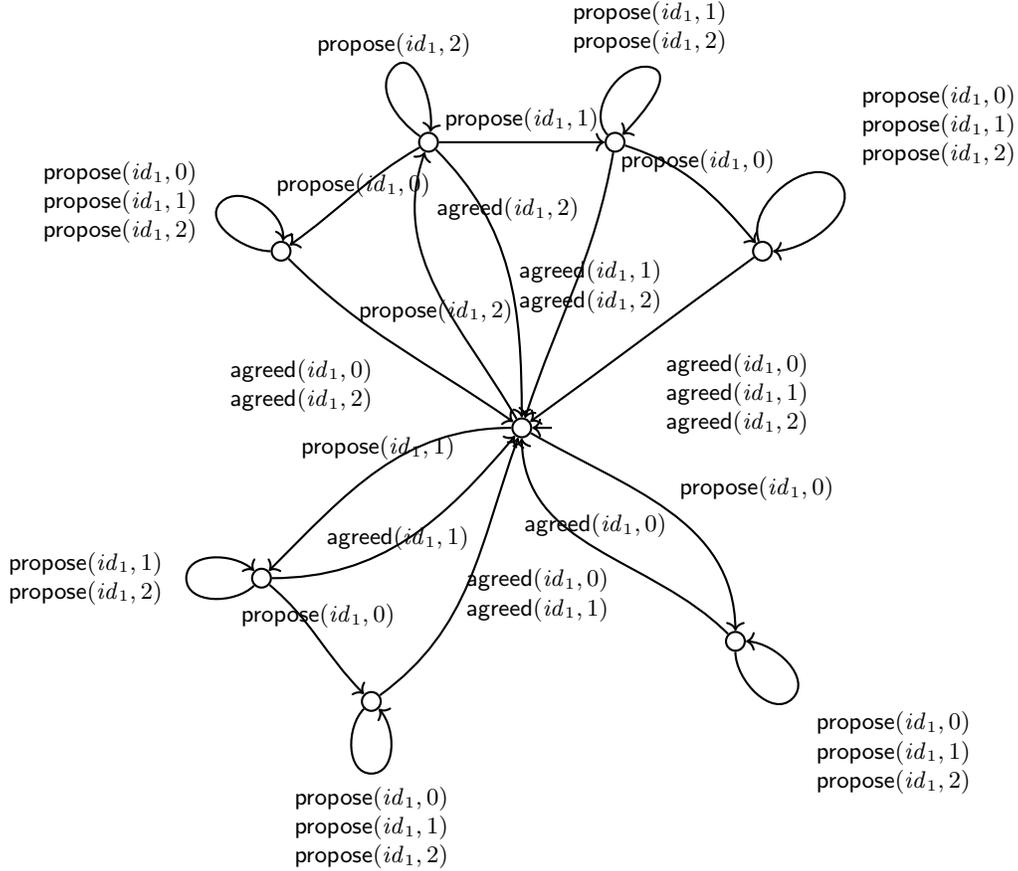
\begin{figure}[t]
\begin{center}
\begin{tikzpicture}
\state{0}{0}{A};
\draw [thick, ->] (0.4,0) -- (A);
\state{3.2}{2.35}{B};
\state{1.24}{3.8}{C};
\state{-1.24}{3.8}{D};
\state{-3.2}{2.35}{E};
\state{-3.46}{-2.00}{F};
\state{-2.00}{-3.64}{G};
\state{ 2.84}{-2.84}{H};

\transition{A}{\action{propose}(\mathit{id}_1,2)}{below}{D}{120}{250};
\transition{D}{\action{agreed}(\mathit{id}_1,2)}{above=15}{A}{-45}{90};
\transition{D}{\action{propose}(\mathit{id}_1,0)}{above=-5}{E}{210}{35};
\transition{E}{\begin{array}{c}\action{agreed}(\mathit{id}_1,0)\\\action{agreed}(\mathit{id}_1,2)\end{array}}{below left}{A}{-45}{145};

\transition{D}{\action{propose}(\mathit{id}_1,1)}{above}{C}{0}{180};
\transition{C}{\begin{array}{c}\action{agreed}(\mathit{id}_1,1)\\\action{agreed}(\mathit{id}_1,2)\end{array}}{right=-30}{A}{-100}{72};

\transition{C}{\action{propose}(\mathit{id}_1,0)}{above}{B}{-18}{130};
\transition{B}{\begin{array}{c}\action{agreed}(\mathit{id}_1,0)\\\action{agreed}(\mathit{id}_1,1)\\\action{agreed}(\mathit{id}_1,2)\end{array}}{below right}{A}{216}{36};

\transition{A}{\action{propose}(\mathit{id}_1,1)}{above}{F}{180}{45};
\transition{F}{\action{agreed}(\mathit{id}_1,1)}{left=-27}{A}{0}{-130};
\transition{F}{\action{propose}(\mathit{id}_1,0)}{above}{G}{-40}{135};
\transition{G}{\begin{array}{c}\action{agreed}(\mathit{id}_1,0)\\\action{agreed}(\mathit{id}_1,1)\end{array}}{below right=-10}{A}{36}{-110};

\transition{A}{\action{propose}(\mathit{id}_1,0)}{above right}{H}{-30}{90};
\transition{H}{\action{agreed}(\mathit{id}_1,0)}{above}{A}{135}{-90};

\selfloop{E}{\begin{array}{c}\action{propose}(\mathit{id}_1,0)\\\action{propose}(\mathit{id}_1,1)\\\action{propose}(\mathit{id}_1,2)\end{array}}
      {left}{-4}{3}{180}{225}{45}{80};

\selfloop{D}{\action{propose}(\mathit{id}_1,2)}{above}{-1.7}{4.8}{145}{225}{45}{80};

\selfloop{C}{\begin{array}{c}\action{propose}(\mathit{id}_1,1)\\\action{propose}(\mathit{id}_1,2)\end{array}}
      {above}{1.7}{4.8}{135}{170}{-10}{45};

\selfloop{B}{\begin{array}{c}\action{propose}(\mathit{id}_1,0)\\\action{propose}(\mathit{id}_1,1)\\\action{propose}(\mathit{id}_1,2)\end{array}}
      {above right}{4.2}{3.3}{110}{135}{-45}{0};

\selfloop{F}{\begin{array}{c}\action{propose}(\mathit{id}_1,1)\\\action{propose}(\mathit{id}_1,2)\end{array}}
      {left}{-4.46}{-2}{225}{-90}{90}{135};

\selfloop{G}{\begin{array}{c}\action{propose}(\mathit{id}_1,0)\\\action{propose}(\mathit{id}_1,1)\\\action{propose}(\mathit{id}_1,2)\end{array}}
      {below}{-2}{-4.6}{225}{180}{0}{-45};

\selfloop{H}{\begin{array}{c}\action{propose}(\mathit{id}_1,0)\\\action{propose}(\mathit{id}_1,1)\\\action{propose}(\mathit{id}_1,2)\end{array}}
      {below right}{3.6}{-3.6}{-90}{225}{45}{0};

\end{tikzpicture}
\end{center}
\caption{The external behaviour for one single party $1$ with three levels of service.}%
\label{figure:threeweak}
\end{figure}

We first focus on the flow of control of process $\expr{P_{env}}$, see the second equation in Figure~\ref{fig:protocol}.
If the two parties have reached a valid decision (which is stored in data parameter $\expr{decision}$), the protocol entity reverts to a mode in which any newly arriving proposals are ignored until the environment is informed about the agreed service level.
As long as no valid decision has been reached, any valid proposal $l$ the entity receives from its environment is processed.
In case the new proposal $l$ is strictly lower than those proposed earlier by the environment (stored in parameter $\expr{mine}$) and not lower than the proposals received so far from the other party (stored in parameter $\expr{theirs}$), the new proposal $l$ is taken into consideration.
In any other case, it is ignored since both parties already communicated a lower service level.
Agreement is reached when the considered proposal $l$ agrees with the least proposal so far received from the other party, \emph{i.e.}, when $l \dataeq \expr{theirs}$; this is communicated to the other party by informing it of decision $l$, and setting Boolean $\expr{hold}$, to indicate that the protocol entity must `suspend' negotiation until also the other party informs the protocol entity that a decision was reached.
If the considered proposal $l$ does not match the lowest service level of the other party, it must decrease the service level considered acceptable for the protocol entity itself.
In this case, the other party is informed of this lower service level.

The Boolean $\expr{hold}$ also dictates the control flow in the process that deals with messages received from the other party, see the fifth equation in Figure~\ref{fig:protocol}.
In case $\expr{hold}$ is high, this means that the protocol entity is currently waiting for the other party to confirm that it has reached a decision.
If this is the case, $\expr{hold}$ becomes low, but, until this happens, all suggested service levels that the protocol entity receives from the other party are ignored.
Whenever negotiation is still in progress, it may be the case that the other party has sent a decision $l$, to which the protocol entity then replies by confirming this decision and storing the decision in parameter $\expr{decision}$; all previously proposed service levels, both from the other party and the protocol entity itself, are reset.
In case the other party has sent a new service level $l$ which was previously proposed through the protocol entity itself, and no service level has yet been decided upon, $l$ is the service level both parties will agree upon.
This is then communicated to the other party.
Whenever $l$ was not previously proposed and is below the recorded minimal proposal (stored in $\expr{theirs}$) received from the other party, the new minimal level is stored.

The service level agreement negotiation protocol, which is the composition of the two protocol entities communicating via the unidirectional channels,
see Figure~\ref{fig:structure}, is described by the following process:
\[
\begin{array}{l}
\keyword{proc}~ \expr{SLAN} = \keyword{allow}(\{\action{in_q}, \action{out_q}, \action{propose},\action{agreed}\},\\
\qquad \qquad \qquad \quad \keyword{comm}(\{ \action{s}|\action{r_q} \to \action{in_q},
             \action{r}|\action{s_q} \to \action{out_q}\}, \\
\qquad \qquad \qquad \qquad \qquad  \expr{P}(\expr{id_1},\expr{id_2},\{\,\},\None,\None,\expr{false}) ~||~\\
\qquad \qquad \qquad \qquad \qquad \expr{Q}(\expr{id_1},[\,]) ~||~Q(\expr{id_2},[\,]) ~||~ \\
\qquad \qquad \qquad \qquad \qquad  \expr{P}(\expr{id_2},\expr{id_1},\{\,\},\None,\None,\expr{false}) ) )
\end{array}
\]

\section{Behavioural reductions}%
\label{sect4}

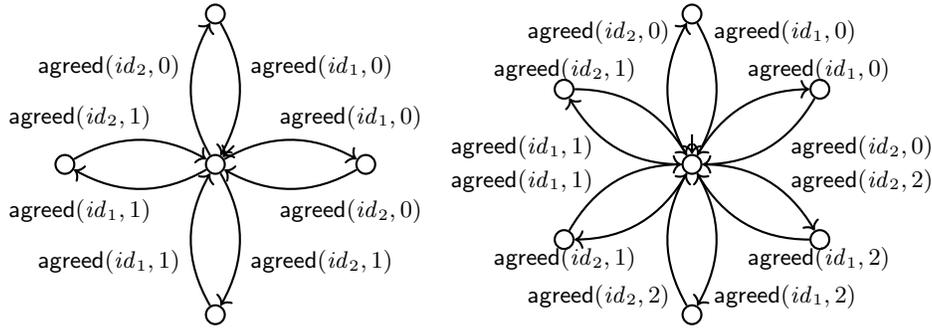
\begin{figure}[t]
\begin{center}
\begin{tikzpicture}
\state{0}{0}{A};
\draw [thick, ->] (0.25,0.25) -- (A);
\state{2}{0}{B};
\state{0}{2}{C};
\state{-2}{0}{D};
\state{0}{-2}{E};

\transition{A}{\hspace*{1.6cm}\action{agreed}(\mathit{id}_1,0)}{above}{B}{30}{150};
\transition{B}{\hspace*{1.6cm}\action{agreed}(\mathit{id}_2,0)}{below}{A}{-150}{-30};

\transition{A}{\action{agreed}(\mathit{id}_2,0)}{above left}{C}{120}{-120};
\transition{C}{\action{agreed}(\mathit{id}_1,0)}{above right}{A}{-60}{60};

\transition{A}{\action{agreed}(\mathit{id}_1,1)\hspace*{1.6cm}}{below}{D}{-150}{-30};
\transition{D}{\action{agreed}(\mathit{id}_2,1)\hspace*{1.6cm}}{above}{A}{30}{150};

\transition{A}{\action{agreed}(\mathit{id}_2,1)}{below right}{E}{-60}{60};
\transition{E}{\action{agreed}(\mathit{id}_1,1)}{below left}{A}{120}{-120};
\end{tikzpicture}
\begin{tikzpicture}
\state{0}{0}{A};
\draw [thick, ->] (0,0.4) -- (A);
\state{1.7}{1}{B};
\state{0}{2}{C};
\state{-1.7}{1}{D};
\state{-1.7}{-1}{E};
\state{0}{-2}{F};
\state{1.7}{-1}{G};

\transition{A}{\hspace*{2cm}\action{agreed}(\mathit{id}_1,0)}{above=5}{B}{60}{180};
\transition{B}{\hspace*{2.5cm}\action{agreed}(\mathit{id}_2,0)}{}{A}{-120}{0};

\transition{A}{\begin{array}{l}\action{agreed}(\mathit{id}_2,0)\\ \\ \\\end{array}}{above left=-10}{C}{120}{-120};
\transition{C}{\begin{array}{l}\action{agreed}(\mathit{id}_1,0)\\ \\ \\\end{array}}{above right=-10}{A}{-60}{60};

\transition{A}{\action{agreed}(\mathit{id}_1,1)\hspace*{2.5cm}}{}{D}{-180}{-60};
\transition{D}{\action{agreed}(\mathit{id}_2,1)\hspace*{2cm}}{above=5}{A}{0}{120};

\transition{A}{\action{agreed}(\mathit{id}_2,1)\hspace*{2cm}}{below=5}{E}{-120}{0};
\transition{E}{\action{agreed}(\mathit{id}_1,1)\hspace*{2.5cm}}{}{A}{60}{180};

\transition{A}{\begin{array}{l}\\ \\\action{agreed}(\mathit{id}_1,2)\end{array}}{below right=-10}{F}{-60}{60};
\transition{F}{\begin{array}{l}\\ \\\action{agreed}(\mathit{id}_2,2)\end{array}}{below left=-10}{A}{120}{-120};

\transition{A}{\hspace*{2.5cm}\action{agreed}(\mathit{id}_2,2)}{}{G}{0}{120};
\transition{G}{\hspace*{2cm}\action{agreed}(\mathit{id}_1,2)}{below=5}{A}{180}{-60};
\end{tikzpicture}

\end{center}
\caption{The external behaviour, reduced with respect to weak trace equivalence, for party $\mathit{id}_1$ with (left) two levels of service, and (right) three levels of service.}%
\label{figure:property3}
\end{figure}
Given the model it is easy to generate the full state space of the protocol containing its full behaviour, although the
sizes of these grow rapidly, see Table~\ref{table:sizes}.
After hiding the internal communications, \ie, writing and reading from queues, and, depending on the use case, also the actions by one of the parties,
we can apply state space minimisation to better understand our external behaviour and the basic properties of the protocol.
The type of behavioural equivalence used when minimising a state space determines which properties are preserved by such a reduction.

Properties (I) and (II) are safety properties on the external traces of the protocol, projected onto the behaviour of a single protocol entity.
Safety properties are preserved when minimising with respect to weak trace equivalence.
Since the properties are only concerned with $\action{propose}$ actions of a fixed party, we can abstract from all $\action{in_q}$ and $\action{out_q}$ actions and all $\action{propose}$ and $\action{agreed}$ actions by the other party.
When we do so, we observe that the minimised state space of our protocol for entity ${\mathit{id}_1}$ when $\mathit{Max}=2$ is exactly as depicted in Figure~\ref{fig:weak_trace_single_party} (see p.~\pageref{fig:weak_trace_single_party}).
The minimised state space of our protocol for entity ${\mathit{id}_1}$ when $\mathit{Max}=3$ is depicted in Figure~\ref{figure:threeweak} (see p.~\pageref{figure:threeweak}).
In both cases, it is easily seen that properties (I) and (II) hold for these two instances.
Obviously, for larger values of $\mathit{Max}$ these pictures become cluttered and therefore rather uninformative.

Property (III) is a safety property that can be visually confirmed by abstracting from all actions of the protocol, except for actions $\action{agreed}$.
Minimising the state space of the protocol using weak trace equivalence leads to small, insightful pictures.
The state space on the left in Figure~\ref{figure:property3} depicts the resulting state space for $\mathit{Max} = 2$, whereas the state space on the right depicts the state space for $\mathit{Max} = 3$.
Both state spaces clearly show that in each round, consisting of two consecutive $\action{agreed}$ actions, the agreed levels of service are identical.

\begin{figure}[p]
\begin{center}
\begin{tikzpicture}
\state{0}{0}{A2};
\draw [thick, ->] (0.4,0.4) -- (A2);
\state{0}{3}{B2};
\state{0}{6}{C2};
\state{0}{9}{D2};
\state{0}{-3}{B2'};
\state{0}{-6}{C2'};
\state{0}{-9}{D2'};

\state{-3}{0.5}{A1};
\state{-3}{-0.5}{A1'};
\state{-6}{0}{A0};
\state{3}{0.5}{A3};
\state{3}{-0.5}{A3'};
\state{6}{0}{A4};

\state{-2}{-3}{B1'};
\state{2}{-3}{B3'};
\state{-2}{3}{B1};
\state{2}{3}{B3};

\state{-2}{-6}{C1'};
\state{-2}{6}{C1};
\state{2}{-6}{C3'};
\state{2}{6}{C3};

\state{-3}{-9}{D1'};
\state{-3}{9}{D1};
\state{3}{-9}{D3'};
\state{3}{9}{D3};

\draw[->]

(A2) edge node[right] {\footnotesize$\action{propose}(\mathit{id}_2,1)$} (B2)
(A2) edge node[right] {\footnotesize$\action{propose}(\mathit{id}_1,1)$} (B2')

(A1) edge node[above] {\footnotesize$\action{agreed}(\mathit{id}_1,1)$}  (A2)
(A1')edge node[below] {\footnotesize$\action{agreed}(\mathit{id}_2,1)$}  (A2)
(A0) edge node[above] {\footnotesize$\action{agreed}(\mathit{id}_2,1)$}  (A1)
(A0) edge node[below] {\footnotesize$\action{agreed}(\mathit{id}_1,1)$}  (A1')

(A4) edge node[below] {\footnotesize$\action{agreed}(\mathit{id}_1,0)$}  (A3')
(A4) edge node[above] {\footnotesize$\action{agreed}(\mathit{id}_2,0)$}  (A3)
(A3) edge node[above] {\footnotesize$\action{agreed}(\mathit{id}_1,0)$}  (A2)
(A3') edge node[below] {\footnotesize$\action{agreed}(\mathit{id}_2,0)$}  (A2)
(A2) edge[bend right] node[sloped,above] {\footnotesize$\action{propose}(\mathit{id}_1,0)$} (D2')
(A2) edge[bend right] node[sloped,above] {\footnotesize$\action{propose}(\mathit{id}_2,0)$} (D2)
(B2) edge[bend right] node[pos=0.4,sloped,below] {\footnotesize$\action{propose}(\mathit{id}_1,0)$} (D2')
(B2') edge[bend right] node[pos=0.4,sloped,above] {\footnotesize$\action{propose}(\mathit{id}_2,0)$} (D2)

(A1') edge node {\footnotesize$\action{propose}(\mathit{id}_1,1)$} (B1')
(A3') edge node {\footnotesize$\action{propose}(\mathit{id}_1,1)$} (B3')
(A3) edge node {\footnotesize$\action{propose}(\mathit{id}_2,1)$} (B3)
(A1) edge node {\footnotesize$\action{propose}(\mathit{id}_2,1)$} (B1)
(B3') edge node[above] {\footnotesize$\action{agreed}(\mathit{id}_2,0)$} (B2')
(B3) edge node[above] {\footnotesize$\action{agreed}(\mathit{id}_1,0)$} (B2)
(B1') edge node[above] {\footnotesize$\action{agreed}(\mathit{id}_2,1)$} (B2')
(B1) edge node[above] {\footnotesize$\action{agreed}(\mathit{id}_1,1)$} (B2)

(B2') edge node {\footnotesize$\action{propose}(\mathit{id}_1,0)$} (C2')
(B2) edge node {\footnotesize$\action{propose}(\mathit{id}_2,0)$} (C2)
(B2') edge node[above] {\footnotesize$\action{propose}(\mathit{id}_2,1)$} (A0)
(B2) edge node[left] {\footnotesize$\action{propose}(\mathit{id}_1,1)$} (A0)
(B3') edge node {\footnotesize$\action{propose}(\mathit{id}_1,0)$} (C3')
(B3) edge node {\footnotesize$\action{propose}(\mathit{id}_2,0)$} (C3)
(B1') edge node {\footnotesize$\action{propose}(\mathit{id}_1,0)$} (C1')
(B1) edge node {\footnotesize$\action{propose}(\mathit{id}_2,0)$} (C1)

(C3') edge node[above] {\footnotesize$\action{agreed}(\mathit{id}_2,0)$} (C2')
(C3) edge node[above] {\footnotesize$\action{agreed}(\mathit{id}_1,0)$} (C2)
(C1') edge node[above] {\footnotesize$\action{agreed}(\mathit{id}_2,1)$} (C2')
(C1) edge node[above] {\footnotesize$\action{agreed}(\mathit{id}_1,1)$} (C2)
(C2') edge node[below,sloped] {\footnotesize$\action{propose}(\mathit{id}_2,1)$} (A0)
(C2) edge node[above,sloped] {\footnotesize$\action{propose}(\mathit{id}_1,1)$} (A0)
(C2') edge node[below,sloped] {\footnotesize$\action{propose}(\mathit{id}_2,0)$} (A4)
(C2) edge node[above,sloped] {\footnotesize$\action{propose}(\mathit{id}_1,0)$} (A4)
(C1') edge node[right] {\footnotesize$\tau$} (D1')
(C1) edge node[right] {\footnotesize$\tau$} (D1)
(C2') edge node[right] {\footnotesize$\tau$} (D2')
(C2) edge node[right] {\footnotesize$\tau$} (D2)
(C3') edge node[right] {\footnotesize$\tau$} (D3')
(C3) edge node[right] {\footnotesize$\tau$} (D3)

(D3') edge node[above] {\footnotesize$\action{agreed}(\mathit{id}_2,0)$} (D2')
(D3) edge node[above] {\footnotesize$\action{agreed}(\mathit{id}_1,0)$} (D2)
(D2') edge node[below,sloped] {\footnotesize$\action{propose}(\mathit{id}_2,0)$} (A4)
(D2) edge node[above,sloped] {\footnotesize$\action{propose}(\mathit{id}_1,0)$} (A4)
(D1') edge node[above] {\footnotesize$\action{agreed}(\mathit{id}_2,1)$} (D2')
(D1) edge node[above] {\footnotesize$\action{agreed}(\mathit{id}_1,1)$} (D2)

(A1') edge[bend right] node[left] {\footnotesize$\action{propose}(\mathit{id}_1,0)$} (D1')
(A1) edge[bend left] node[left] {\footnotesize$\action{propose}(\mathit{id}_2,0)$} (D1)
(A3') edge[bend left] node[right] {\footnotesize$\action{propose}(\mathit{id}_1,0)$} (D3')
(A3) edge[bend right] node[right] {\footnotesize$\action{propose}(\mathit{id}_2,0)$} (D3)
;


\end{tikzpicture}

\end{center}
\caption{The external behaviour modulo divergence-preserving branching bisimulation for $\mathit{Max}=2$.}%
\label{fig:branchbis}
\end{figure}
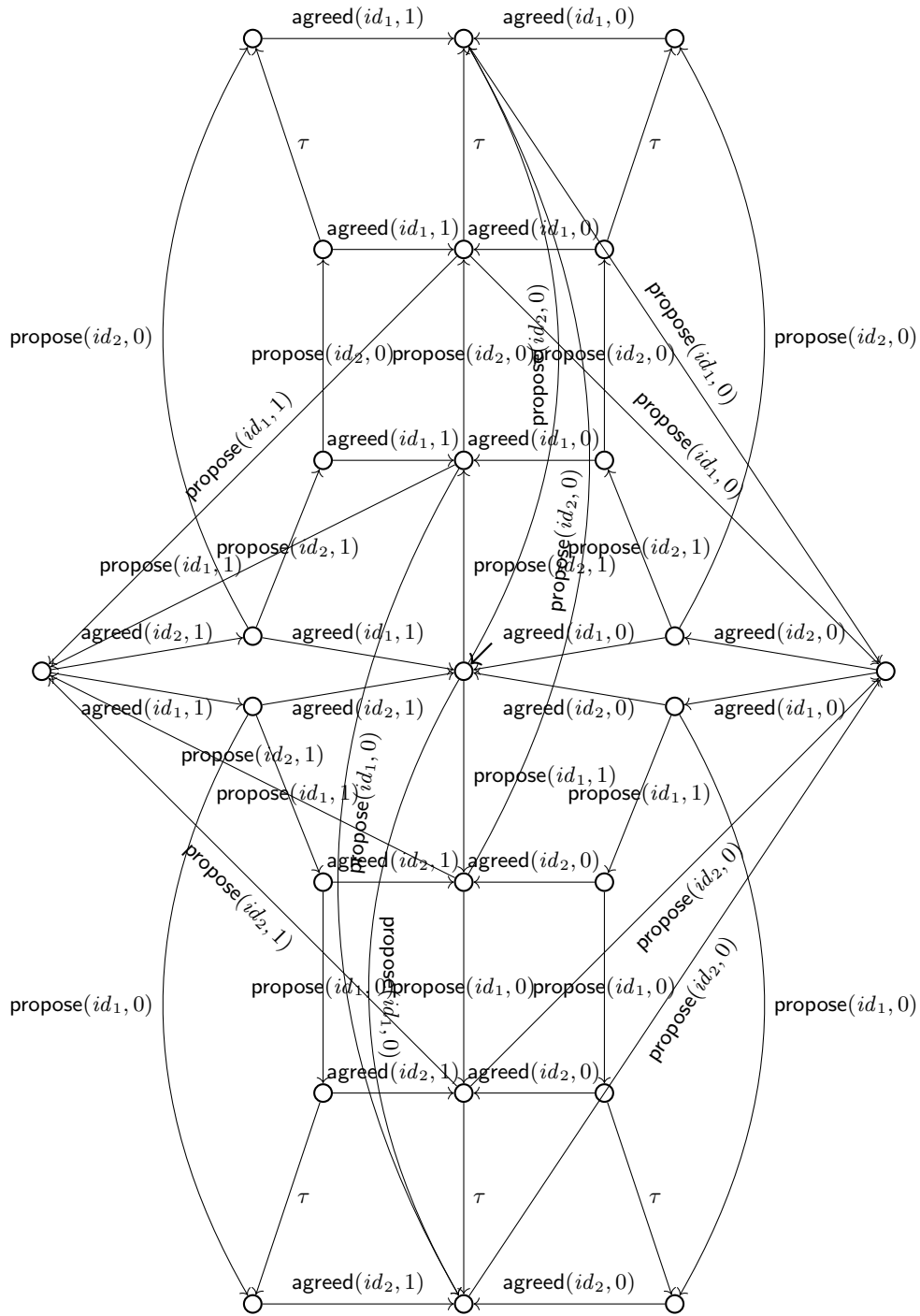

In case of property (IV), which is a liveness property, a behavioural equivalence relation that is more discriminating than weak trace equivalence is required for minimising the state space to preserve the property.
We use divergence-preserving branching bisimilarity~\cite{DBLP:journals/fuin/GlabbeekLT09} to reduce the state space, abstracting only from actions $\action{in_q}$ and $\action{out_q}$;
see Figure~\ref{fig:branchbis} for the resulting state space for $\mathit{Max}=2$.
All self-loops have been omitted for the sake of readability. This means that if there is no explicit outgoing transition
in a state labelled with a $\action{propose}$ action, there is a self-loop in that state with such actions. This reduced state space has
$25$ states and $126$ transitions, among which $6$ $\tau$-transitions that correspond to reading a message from a queue, inhibiting
certain service level agreements, and which are therefore present in the state space even after minimisation.

It may be clear that property (IV) is hard to confirm visually even for such small values of $\mathit{Max}$.
The initial state is the state in the middle. The state at the right corresponds to the system agreeing to a service level $0$, and
the state at the left corresponds to agreeing to service level $1$. The central column (up and down) corresponds to the situation where both protocol
entities are in the same round. In the columns at the left and the right, one of the parties still has to report the conclusion of the
previous round. In the right upper quadrant party ${\mathit{id}_1}$ has still to report agreement on service level $0$, in the left upper quadrant
it has still to report service level $1$. In the two lower quadrants protocol entity ${\mathit{id}_2}$ still has to report the service level agreed upon.

While moving further up or down, the protocol tends to settle on a lower service level. This is best seen by the transitions
going to the rightmost and leftmost states. In the uppermost and lowermost state, transitions to the leftmost state are no longer possible,
where the quality of service is settled on level $1$.
Confirming property (IV) on this transition system is a gruesome task, but feasible with a little perseverance. It is
rather undoable to verify requirement (IV) for larger values of $\mathit{Max}$. \medskip

The state space of the protocol also permits to inspect all conceivable contents of the queues.
By design, the protocol only submits relevant proposals from a party to the other party, and, therefore, implicitly puts an upper bound on the occupancy of the queues.
While our initial assumption was that each
queue would contain at most one $\mathit{decide}$  message and $\mathit{Max}$ $\mathit{inform}$ messages, this turns out not to be true.
In fact, for the case when $\mathit{Max} = 2$, up to seven simultaneous messages can be present in the queues, as illustrated by the queue contents depicted below.
\[\begin{array}{l}
[\mathit{inform}(0), \mathit{inform}(1), \mathit{decide}(1), \mathit{inform}(0), \mathit{inform}(1), \mathit{decide}(1), \mathit{inform}(0)]
\end{array}\]
Note that the leftmost message is the most recent one.
The queue corresponds to the situation where one process confirms the decision of two rounds earlier. The rightmost $\mathit{decide}(1)$,
marks a decision for two rounds prior.
The leftmost $\mathit{decide}(1)$ indicates that also the previous round settled on service level $1$; the $\mathit{inform}$ messages in between the two $\mathit{decide}$ messages are those leading to the decision.
\begin{figure}[h]\centering
	\begin{tikzpicture}
		\node[rectangle,minimum width=1.5cm, minimum height=.5cm,draw] (id1) at (0,.3) {\small $\mathit{id}_1$};
		\node[rectangle,minimum width=1.5cm, minimum height=.5cm,draw] (Q1) at (3,.3) {\small \it queue 1};
		\node[rectangle,minimum width=1.5cm, minimum height=.5cm,draw] (Q2) at (6,.3) {\small \it queue 2};
		\node[rectangle,minimum width=1.5cm, minimum height=.5cm,draw] (id2) at (9,.3) {\small $\mathit{id}_2$};

		\node[draw=none,below of=id1, yshift=-11.5cm] (id1') {};
		\node[draw=none,below of=Q1, yshift=-11.5cm] (Q1') {};
		\node[draw=none,below of=Q2, yshift=-11.5cm] (Q2') {};
		\node[draw=none,below of=id2, yshift=-11.5cm] (id2') {};

		\draw (id1) -- (id1');
		\draw (id2) -- (id2');
		\draw (Q1) -- (Q1');
		\draw (Q2) -- (Q2');
		\draw[->] (-2,-1) -- node[midway,above] {\scriptsize $\action{propose}(1)$} (0,-1); 
		\draw[->] (0,-1.5) -- node[midway,above] {\scriptsize $\mathit{inform(1)}$} (3,-1.5); 
		\draw[->] (3,-2) -- node[near end,above] {\scriptsize $\mathit{inform(1)}$} (9,-2); 
		\draw[->] (-2,-2) -- node[midway,above] {\scriptsize $\action{propose}(0)$} (0,-2); 
		\draw[->] (0,-2.5) -- node[midway,above] {\scriptsize $\mathit{inform(0)}$} (3,-2.5); 
		\draw[->] (11.0,-2.5) -- node[midway,above] {\scriptsize $\action{propose}(1)$} (9,-2.5); 
		\draw[->] (9,-3) -- node[midway,above] {\scriptsize $\mathit{decide(1)}$} (6,-3); 
		\draw[->] (9,-3.5) -- node[midway,above] {\scriptsize $\action{agreed}(1)$} (11.0,-3.5); 
		\draw[->] (11.0,-4) -- node[midway,above] {\scriptsize $\action{propose}(1)$} (9,-4); 
		\draw[->] (9,-4.5) -- node[midway,above] {\scriptsize $\mathit{inform(1)}$} (6,-4.5); 
		\draw[->] (6,-5) -- node[near end,above] {\scriptsize $\mathit{decide(1)}$} (0,-5); 
		\draw[->] (0,-5.5) -- node[midway,above] {\scriptsize $\mathit{decide(1)}$} (3,-5.5); 
		\draw[->] (0,-6) -- node[midway,above] {\scriptsize $\action{agreed}(1)$} (-2,-6); 
		\draw[->] (-2,-6.5) -- node[midway,above] {\scriptsize $\action{propose}(1)$} (0,-6.5); 
		\draw[->] (0,-7) -- node[midway,above] {\scriptsize $\mathit{inform(1)}$} (3,-7); 
		\draw[->] (-2,-7.5) -- node[midway,above] {\scriptsize $\action{propose}(0)$} (0,-7.5); 
		\draw[->] (0,-8) -- node[midway,above] {\scriptsize $\mathit{inform(0)}$} (3,-8); 
		\draw[->] (6,-8.5) -- node[near end,above] {\scriptsize $\mathit{inform(1)}$} (0,-8.5); 
		\draw[->] (3,-9) -- node[midway,above] {\scriptsize $\mathit{decide(1)}$} (0,-9); 
		\draw[->] (0,-9.5) -- node[midway,above] {\scriptsize $\action{agreed}(1)$} (-2,-9.5); 
		\draw[->] (-2,-10) -- node[midway,above] {\scriptsize $\action{propose}(1)$} (0,-10); 
		\draw[->] (0,-10.5) -- node[midway,above] {\scriptsize $\mathit{inform(1)}$} (3,-10.5); 
		\draw[->] (-2,-11) -- node[midway,above] {\scriptsize $\action{propose}(0)$} (0,-11); 
		\draw[->] (0,-11.5) -- node[midway,above] {\scriptsize $\mathit{inform(0)}$} (3,-11.5); 
	\end{tikzpicture}
	\caption{Sequence diagram illustrating that for $\mathit{Max}=2$, there can be seven messages present in a queue.}%
	\label{fig:sequence_diagram}

\end{figure}
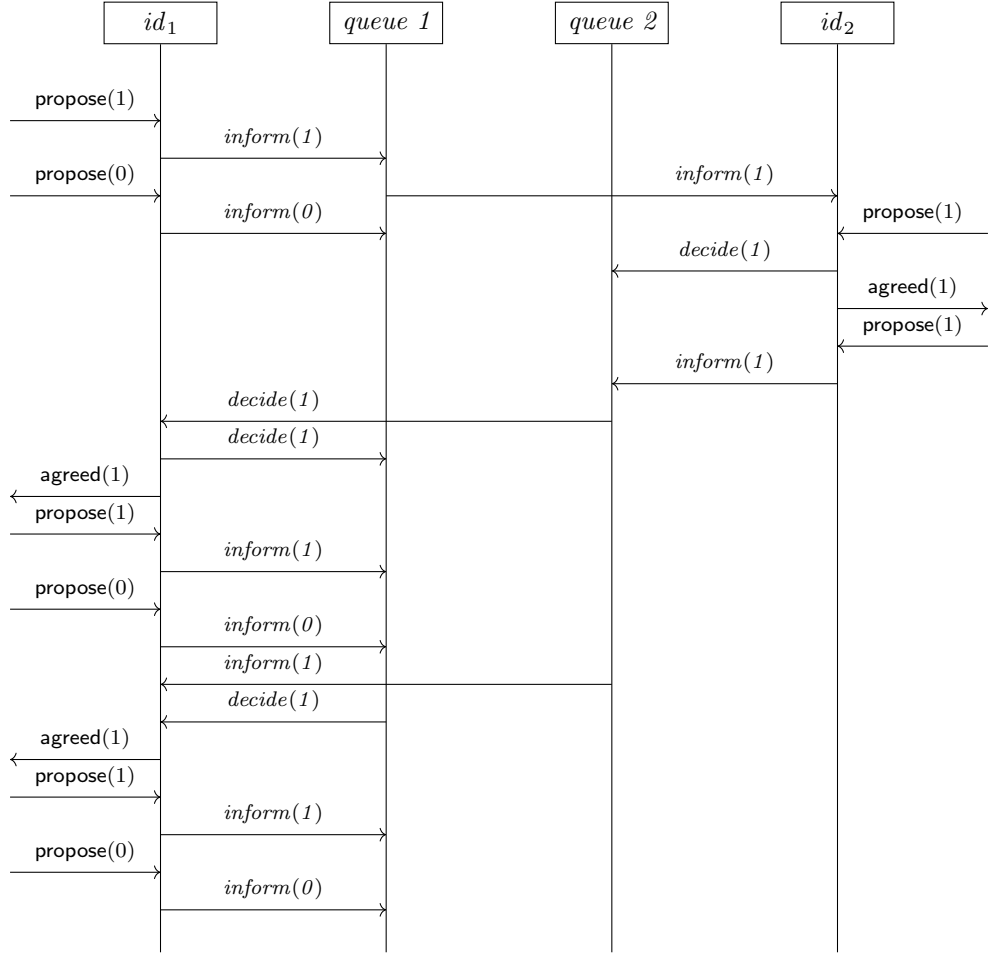

The sequence diagram of Figure~\ref{fig:sequence_diagram} depicts executions of the protocol that lead to this particular queue contents.
We observe that the $\mathit{inform}$ messages in the queue are strictly decreasing in each round, so there can be
at most $\mathit{Max}$ consecutive such messages for a single round.
The data for three rounds can be simultaneously in the queue. This is nicely illustrated in
Figure~\ref{fig:sequence_diagram}. For the oldest round, agreement has been reached and reported. The oldest
$\mathit{inform}$ message, communicating a proposal for the highest service level, has been consumed by $\mathit{id}_2$.
This is needed to
reach a decision. For the second round process $\mathit{id}_2$ sends one $\mathit{inform}$ message, which is sufficient
for $\mathit{id}_1$ to reach a decision. But before reaching a decision, $\mathit{id}_1$ can already send all,
\ie, $\mathit{Max}$ distinct $\mathit{inform}$ messages in the queue, which is then followed by the decision. Although
process $\mathit{id}_2$ did not read any of these messages, and reach a decision in this round, process
$\mathit{id}_1$ can already be fully engaged in the next round, filling the queue again with a whole sequence
of $\mathit{Max}$ $\action{inform}$ messages.

Summarising, for the oldest
round there are $\mathit{Max}-1$ $\mathit{inform}$ messages, as the receiving party has to have read one $\mathit{inform}$ message to come
to an agreement, and one $\mathit{decide}$ message indicating a decision. These are $\mathit{Max}$ messages in total. For the
second round there can be $\mathit{Max}$ distinct $\mathit{inform}$ messages and one $\mathit{decide}$ message,
totalling $\mathit{Max}+1$. For the third round there can again be $\mathit{Max}$ distinct $\mathit{inform}$ messages, and no
agreement can have been reached, so there is no $\mathit{decide}$ message.
If we add up all these messages, we find that there can be at most $3\mathit{Max}+1$ messages in one queue simultaneously.


\section{Formal correctness requirements}%
\label{sect5}
In order to convince ourselves that the protocol is correct for larger values of $\mathit{Max}$, we resort to model checking and formalise the requirements of Section~\ref{sec:requirements} in modal logic.
First note that all requirements are essentially requirements on the execution paths of the protocol or, in case of requirement (IV), \emph{spines}: execution paths augmented with information concerning enabled actions in states along such paths; moreover,
this last property assumes some form of fair behaviour of the two involved parties.
Second, note that all properties reason about (sets of) service levels proposed during a round.
The former requires a logic capable of reasoning about paths and their branches (\ie, a \emph{branching time logic}); the second benefits from a logic that supports variables for memorising and updating these (sets of) proposals and reasoning about these.
Since our model is action-based, we therefore use the modal mu-calculus with data~\cite{DBLP:conf/amast/GrooteM98,DBLP:journals/scp/GrooteW05,DBLP:journals/iandc/PloegerWW11} to express the properties.

\subsection{A note on the modal mu-calculus with data}%
\label{sec:mu-calc}

For a detailed exposition of the syntax and semantics of this logic, including a large number of examples, we refer to~\cite{GM14,GKLVW20}.
We here briefly recall the basic language constructs.
\emph{Boolean expressions} and logical operators, such as conjunction, disjunction, recursion variables and first-order quantification, are the basic building blocks of the modal mu-calculus with data.
The modal operators $\langle \alpha \rangle \phi$ and  $[\alpha]\phi$ allow for reasoning about the transitions of a transition system: formula $\langle \alpha \rangle \phi$ holds of a state if it it has an $\alpha$-successor satisfying $\phi$, whereas $[\alpha]\phi$ holds of a state if all its $\alpha$-successors satisfy $\alpha$.
Of particular note here is that $\alpha$ characterises a set of actions: $a(e)$ denotes the singleton set consisting of the parameterised action $a(e)$; the expressions $\alpha \cup \beta$ and $\alpha \cap \beta$ denote set union and set intersection, respectively; for a set described by $\alpha$, $\bar{\alpha}$ denotes its complement; and $\exists d{:}D.\alpha$ denotes the union of all collections obtained by instantiating $d$ in $\alpha$. For instance, the formula $\true$ describes the set of all actions, whereas $\exists n{:}\mathbb{N}. a(n)$ describes the set of actions $\{a(i) \mid i \in \mathbb{N} \}$.
The modal mu-calculus derives its expressive power from the fixed point operators $\mu$ and $\nu$, which, informally, can be understood as tools to specify finite recursion (\eg, liveness properties) and potentially infinite recursion (\eg, safety properties and invariants).
For instance, the absence of deadlock is formalised by $\nu X. ([\true]X \wedge \langle \true \rangle \true)$,
whereas the formula $\nu X. ([\true]X \wedge \forall l{:}\Nat. [\exists \mathit{id}{:}\mathit{ID}.\action{propose}(id,l)](l \in \expr{Levels}) )$ expresses that any process can only propose valid values.
In a setting with data, the parameterisation of fixed points enable one to recall, store and update information (data) in recursions.

\subsection{Requirement (I)}
The first requirement states that any service level a party agrees to must have been proposed by that party in that round.
We model this property by maintaining a finite set $\mathit{proposed}$ of service levels that have been proposed by party $\mathit{id}$
in one round. Whenever an action $\action{propose}(\mathit{id},l)$ takes place the proposed service level $l$ is added to the set
$\mathit{proposed}$. Whenever, an action $\action{agreed}(\mathit{id},l)$ takes place, the agreed level $l$ must be part of this set, and as
a new round is started the set $\mathit{proposed}$ must be reset. The last line of the modal formula asserts that the requirement holds
with $\mathit{proposed}$ unmodified when any action, other
than $\action{propose}$ or $\action{agreed}$ occurs.
\[\begin{array}{l}
\forall \mathit{id}{:}\mathit{ID}.
         \nu X(\mathit{proposed}:\mathit{Set}(\Nat)=\emptyset).\\
\hspace*{2.0cm}        (\forall l{:}\Nat. [\action{propose}(\mathit{id},l)]X(\mathit{proposed}\cup\{l\})) \wedge{}\\
\hspace*{2.0cm}        (\forall l{:}\Nat. [\action{agreed}(\mathit{id},l)](l \in \mathit{proposed}{} \wedge X(\emptyset))) \wedge{}\\
\hspace*{2.0cm}        ([\overline{\exists l{:}\Nat.
                        \action{propose}(\mathit{id},l){\cup} \action{agreed}(\mathit{id},l)}]X(\mathit{proposed})).
\end{array}\]
We remark that the parameterisation of fixed points is not strictly needed; \ie, the property can also be formalised by a more `classical' formula with nested fixed points. But we find that using data in modal mu-calculus formulas
makes the formulas easier to understand as data parameters often match natural concepts.

\subsection{Requirement (II)}
The second requirement expresses that any party $\mathit{id}$ proposing a service level that is higher than the lowest service level already proposed by the party cannot supersede that lower service level. The service level that is agreed upon is therefore never of a level that is rejected in the negotiation process.
This formula is similar in spirit to the previous formula, except that, in a round, proposed service levels are only added to the set
$\mathit{proposed}$ if they are lower than the current minimum, and an additional set $\mathit{rejected}$ is maintained to keep track of all the rejected levels that have been proposed.
Note that this set does not simply consist of the set of levels that are larger than the least level proposed so far, since a service level higher than this least level may still be under consideration if it has been proposed at an earlier stage.
We can therefore only reject proposals for service levels that have not previously been proposed and that are not lower than the current minimum; these are exactly those service levels added to a set $\mathit{rejected}$.
The formula then asserts that no service level of the set $\mathit{rejected}$ can be agreed upon in a round.
\[\begin{array}{l}
\forall \mathit{id}{:}\mathit{ID}. \nu X(\mathit{proposed}:\mathit{Set}(\Nat)=\emptyset, \mathit{rejected}:\mathit{Set}(\Nat) = \emptyset). \\
\hspace*{0.75cm} (\forall l{:}\Nat. [\action{propose}(\mathit{id},l)](\, (l {\notin} \mathit{proposed} \wedge l {>} \mathit{min}(\mathit{proposed}){} \wedge X(\mathit{proposed},\mathit{rejected} {\cup} \{l\}) ) \vee {} \\
\hspace*{4.75cm}                 (l {\in} \mathit{proposed}{} \wedge X(\mathit{proposed}, \mathit{rejected}) ) \vee {} \\
\hspace*{4.75cm}                 (l {\leq} \mathit{min}(\mathit{proposed}){} \wedge X(\mathit{proposed} {\cup} \{l\}, \mathit{rejected}) )\,) \,) \wedge {} \\
\hspace*{0.75cm} (\forall l{:}\Nat. [\action{agreed}(\mathit{id},l)](l {\notin} \mathit{rejected} \wedge X(\emptyset,\emptyset)) \,) \wedge{}\\
\hspace*{0.75cm} ([\overline{\exists l{:}\Nat.\action{propose}(\mathit{id},l){\cup}\action{agreed}(\mathit{id},l)}]X(\mathit{proposed},\mathit{rejected})).
\end{array}\]

\subsection{Requirement (III)}
Requirement (III) states that if a service level $l$ is agreed upon during a round, then both parties agree on the same service level.
As the rounds are asynchronous, it can be the case that in the current round, no service level has been agreed upon yet (\ie, no $\action{agreed}(\mathit{id},q)$ action for the current round has happened),
or one of the parties has already reported on a service level that is agreed upon (\ie, some $\action{agreed}(\mathit{id},l)$ has been reported in the current round).
To account for this asynchronicity, we therefore use parameters
$a_1$ and $a_2$ of type $\Nat$ to store the service level agreed upon in a round by party $\mathit{id}_1$, resp.~$\mathit{id}_2$, and that contain
$\None$ if no such value for the current round has been announced yet. Whenever $\action{agreed}(\mathit{id},l)$ is reported by party $\mathit{id}$,
the announced value must match that of the other party if it has been announced (\ie, if the associated parameter is different from $\None$).
\[\begin{array}{l}
\nu X(a_1:\Nat=\None,a_2:\Nat=\None).  \\
\hspace*{0.5cm}    (\forall l{:}\Nat. [\action{agreed}({\mathit{id}_1},l)](  ( a_2 {\ndataeq} \None \wedge l {\dataeq} a_2 \wedge X(\None,\None) )
				  \vee ( a_1 {\dataeq} a_2 {\dataeq} \None \wedge X(l,a_2))  )) \wedge{}\\
\hspace*{0.5cm}   (\forall l{:}\Nat. [\action{agreed}({\mathit{id}_2},l)](  ( a_1 {\ndataeq} \None \wedge l {\dataeq} a_1 \wedge
                     X(\None,\None) )
				  \vee (a_1 {\dataeq} a_2 {\dataeq} \None \wedge X(a_1,l)) ))  \wedge{} \\
\hspace*{0.5cm}   ([\overline{\exists \mathit{id}{:}\mathit{ID}, l{:}\Nat.\action{agreed}(\mathit{id},l)}]X(a_1,a_2)).

\end{array}\]
\subsection{Requirement (IV)}
The fourth requirement states that whenever matching proposals have been made, inevitably both parties will be informed
about an agreed quality of service.
Since the protocol always returns to a state in which it is able to accept new proposals, the agreement can be postponed indefinitely by proposing service levels that will nevertheless be ignored.
Once the protocol is in a state in which an agreement can be sent, however, it remains in such a state until it has actually been sent.

We note that both parties may potentially agree to a service level that is below their currently matching proposals.
This is the case if one party submits a lower level after earlier matching proposals have been submitted, or if one of the parties also submitted proposals for lower service levels.
For instance, if party $\mathit{id}_1$ first proposes service level 2, followed by 1, and party $\mathit{id}_2$ then responds by submitting service level 2, it may still be the case that party $\mathit{id}_1$ lowers the desired service level even further by submitting 0.
In such a case agreement might then only be possible if $\mathit{id}_2$ submits level 0 as well, even though level 0 was not among the service levels at the time both parties first submitted a matching proposal.

We formulate this property using two sets of proposed quality of services and whenever the intersection of these sets is not
empty, it is required that $\action{agreed}$ actions must take place.

However, due to asynchronicity, both protocol entities
may be in different rounds. Therefore, we keep lists of finite sets of proposed quality of services $q_1$ for process $1$ and
$q_2$ for process $2$. The first element in the list
contains the set of quality of services proposed in the current round that is not yet finished by both parties. It can be
that one of the processes is already collecting proposals for the next round. In that case its list contains two elements and
newly proposed quality of services are inserted in the second set.
Note that the sum of the lengths of the lists of $q_1$ and $q_2$ can be $2$ or $3$.

In order to understand the formula after the condition $\mathit{head}(q_1) \ndataeq \None \wedge \mathit{head}(q_2) \ndataeq \None$,
it is useful to understand the following simpler formula:
\[
    \mu Z.\nu Y.([\overline{a{\cup}b}]Z \wedge [b]Y \wedge \langle\overline{b}\rangle\true).
\]
This formula says that action $a$ will become enabled within a finite number of non-$b$ steps but any $b$-action may
indefinitely postpone this moment; until this moment, non-$b$ steps remain enabled and following these brings the inevitability of $a$ closer.

In our particular setting, the set of actions represented by $a$ and $b$ is not static but dynamic; the reason is that
before agreement becomes persistently enabled, the parties may still be negotiating and the final agreement is therefore not yet
settled in all circumstances.
Our formula therefore asserts that agreement becomes inevitable---if it is not decided already---when
one party proposes to lower the service level to one that matches the other party's minimum level.
We keep track of these minimum proposed levels by means of two parameters of the $Z$ and $Y$ fixed point variables, and update these
parameters with every occurrence of a $\action{propose}$ action.

In the formulation below, the action $a$ takes the role of the $\action{agreed}(\mathit{id},q)$ action (for universally quantified $\mathit{id}$).
The set of actions represented by $b$ within the must-modality is replaced by the set of all $\action{propose}$ actions, and the set of $b$ actions within
the may-modality is replaced by all $\action{propose}$ actions, except for the possibly necessary proposal matching the minimum of the proposals of the other
party.
The outermost fixed point $X$ is used to assert that the inevitability of agreements is an invariant. Both parameters $q_1$ and $q_2$ are necessary to
keep track of the current minimal proposals made by the two parties in a round. Due to asynchronicity, we may need information about the minimal level
proposed in a previous round. To this end, we use lists of service levels.

\[\begin{array}{l}
  (\nu X(q_1{:}\mathit{List}(\Nat)=[\None], q_2{:}\mathit{List}(\Nat)=[\None]).\\
\hspace{0.5cm}        (\forall l{:}\Nat.[\action{propose}({\mathit{id}_1},l)]X(\mathit{rtail}(q_1){\triangleleft}\expr{min}(\expr{rhead}(q_1),l),q_2)) \wedge{}\\
\hspace{0.5cm}        (\forall l{:}\Nat.[\action{propose}({\mathit{id}_2},l)]X(q_1,\mathit{rtail}(q_2){\triangleleft}\expr{min}(\expr{rhead}(q_2),l))) \wedge{}\\
\hspace{0.5cm}        ([\exists l{:}\Nat.\action{agreed}({\mathit{id}_1},l)]( ( |q_2| {\ndataeq} 1 {\wedge}X(\mathit{tail}(q_1){\triangleleft}\None,
                          \mathit{tail}(q_2)) )\vee( |q_2| {\dataeq} 1 {\wedge} X(q_1{\triangleleft}\None,q_2)) ) ) \wedge{}\\
\hspace{0.5cm}        ([\exists l{:}\Nat.\action{agreed}({\mathit{id}_2},l)](( |q_1| {\ndataeq} 1 {\wedge}
                                 X(\mathit{tail}(q_1),\mathit{tail}(q_2){\triangleleft}\None) )
                                         \vee( |q_1| {\dataeq}1 {\wedge} X(q_1,q_2{\triangleleft}\None)) ) ) \wedge{}\\
\hspace{0.5cm}        ([\overline{\exists \mathit{id}{:}\mathit{ID}, l{:}\Nat.\action{propose}(\mathit{id},l){\cup}
                               \action{agreed}(\mathit{id},l)}]X(q_1,q_2)) \wedge{}\\
\hspace{0.5cm}        ((\expr{head}(q_1) {\ndataeq} \None \wedge \expr{head}(q_2){\ndataeq}\None) \Rightarrow{}\\
\hspace{1.0cm}          \forall \mathit{id}{:}\mathit{ID}. ( (\mathit{id} \dataeq {\mathit{id}_1} \wedge |q_1| {\dataeq} 1) \vee (\mathit{id} \dataeq {\mathit{id}_2} \wedge |q_2| {\dataeq} 1) ) \Rightarrow{} \\
\hspace{1.5cm}             \mu Z(m_1'{:}\Nat = \mathit{head}(q_1), m_2'{:}\Nat = \mathit{head}(q_2)).\\
\hspace{2.0cm}                 \nu Y(m_1{:}\Nat = m_1', m_2{:}\Nat = m_2').\\
\hspace{2.7cm}                      ( [\overline{\exists l{:}\Nat.\action{agreed}(\mathit{id},l){\cup}
                                        \exists \mathit{id}'{:}\mathit{ID}.\action{propose}(\mathit{id}',l)}]Z(m_1,m_2) \wedge{}\\
\hspace{2.7cm}                           (\forall l{:}\Nat. [\action{propose}(\mathit{id}_1,l)] Y(\expr{min}(m_1,l),m_2) ) \wedge{}\\
\hspace{2.7cm}                           (\forall l{:}\Nat. [\action{propose}(\mathit{id}_2,l)] Y(m_1,\expr{min}(m_2,l)) ) \wedge{}\\
\hspace{2.7cm}                          (\langle\,\overline{\exists l{:}\Nat.( (l \ndataeq m_2 \cup m_1 \leq m_2 ) \cap
                                       \action{propose}({\mathit{id}_1},l))} \cap {} \\
\hspace{3.0cm}
                                          \overline{\exists l{:}\Nat.( (l \ndataeq m_1 \cup m_2 \leq m_1) \cap
                                       \action{propose}({\mathit{id}_2},l))}\,\rangle\true) ) )

\end{array}\]
\subsection{Verification}
The four properties above have been verified for $\mathit{Max}$ ranging from $1$ up-to and including $5$ using the mCRL2 toolset~\cite{BunteGKLNVWWW19}.
All four properties hold for all these five instances.
We note that also the two properties stated in Section~\ref{sec:mu-calc} hold for these five instances.
There are several ways of performing such a verification.
Due to the relatively elegant external behaviour of the protocol, it turns out that it is most fruitful to first generate the
state space for each instance, reduce these modulo divergence-preserving branching bisimulation~\cite{DBLP:journals/fuin/GlabbeekLT09}, which preserves the validity of the formulas,
and verify the modal formulas on the reduced state spaces. In Table~\ref{table:sizes} we indicate the sizes of the generated state space.
Note that these grow quickly with increasing $\mathit{Max}$.
Suffices `k', `M' and `G' stand for $10^3$, $10^6$ and $10^9$, respectively.
The column labelled `mod dpbb' depicts the sizes after reduction modulo divergence-preserving branching
bisimulation. Note the very substantial reductions.
\begin{table}[h]
\begin{center}
\begin{tabular}{rrrrr}
$\mathit{Max}$&\#states&\#transitions&\#states mod dpbb&\#transitions mod dpbb\\
\toprule
$1$&$233$&$746$&$8$&$22$\\
$2$&$7.9$k&$38$k&$25$&$126$\\
$3$&$239$k&$1.6$M&$66$&$482$\\
$4$&$6.7$M&$57$M&$163$&$1550$\\
$5$&$176$M&$1.8$G&$388$&$4.5$k\\
$6$&$4.4$G&$541$G&$901$&$12.3$k\\
\end{tabular}
\end{center}
\caption{Sizes of the state spaces for various values of $\mathit{Max}$}%
\label{table:sizes}
\end{table}

\section{Conclusions and future work}%
\label{sect6}

We designed a symmetric protocol for repeatedly negotiating a desired quality of service level between two parties.
We spent considerable effort in ensuring that the state space of the external behaviour of our protocol is as `small' and `insightful' as possible. This is to say that we did not manage to obtain smaller
state spaces without violating the listed desirable properties or by inadvertently introducing deadlocks.
It is, however, reasonably straightforward to make changes that yield more complex state spaces (also after reduction) and that even satisfy all four modal formulas expressing the correctness of our protocol.
We found that due to the concise external behaviour, we could verify the modal formulas for very substantial state spaces as these state
spaces reduce very considerably. This is in line with one of the design rules for verifiable systems as mentioned in~\cite{DBLP:journals/stvr/GrooteKO15}.

There are a few directions for future work. For instance, the
protocol can be generalised to serve $n$ parties.  Furthermore, it
is an open question how to prove the validity of our modal formulas
for our protocol for arbitrary value for $\mathit{Max}$. The complexity of
the protocol is such that any convincing correctness argument requires
mechanical proof checking or full automation.
One option
can be to develop proof methods for parameterised Boolean equation
systems~\cite{GrooteW05}, inspired by the \emph{cones and foci}
method~\cite{DBLP:journals/jlp/GrooteS01,DBLP:conf/fossacs/FokkinkP03},
which is very effective for proving an implementation behaviourally
equivalent to a specification for $n$ processes and infinite data
domains, and which can be mechanised in a proof checker~\cite{FokkinkPP06}.
Other options, which, in theory, work for properties (I)--(III), can be to automate 
proof methods based on invariants~\cite{OrzanW10} or
techniques such as abstract interpretation~\cite{CranenGWW15}.
However, for property (IV), it is conceivable that any such technique will fail
due to the presence of the least fixed point operator, see~\eg~\cite{CranenGWW15},
thus calling for even stronger proof methods.

\section*{Acknowledgements}
Thanks go to the members of the chuck-swap software redesign team of ASML, especially Dennis Verhaegh, Ernesto Romero Sahagun and Stefan Rijkers to
introduce us to the problem of reaching agreement for two service levels with two symmetric parties, making us aware how tricky
it is to come up with a correct protocol to solve this problem, and for providing feedback on earlier drafts of this paper.
Furthermore, thanks go to the reviewers for their valuable feedback and suggestions for improvement.

\bibliographystyle{alpha}

\newcommand{\etalchar}[1]{$^{#1}$}

\end{document}